\documentclass[crop]{CSL}

\jname{International Journal of Astrobiology}%

\begin{document}

\supertitle{Research Paper}

\title[The Fermi paradox: Impact of astrophysical processes and dynamical evolution]{The Fermi paradox: Impact of astrophysical processes and dynamical evolution}

\author[Schleicher \& Bovino]{Dominik R.G. Schleicher$^{1}$ and Stefano Bovino$^{1,2}$}

\address{\add{1}{Departamento de Astronom\'ia, Facultad Ciencias F\'isicas y Matem\'aticas, Universidad de Concepci\'on, Av. Esteban Iturra s/n Barrio Universitario, Casilla 160-C, Concepci\'on, Chile}, \add{2}{INAF - Istituto di Radioastronomia, Via Gobetti 101, I-40129 Bologna, Italy;}}

\corres{\name{D.R.G. Schleicher} \email{dschleicher@astro-udec.cl}}

\begin{abstract}
The Fermi paradox has given rise to various attempts to explain why no evidence of extraterrestrial civilisations was found so far on Earth and in our Solar System. Here, we present a dynamical model for the development of such civilisations, which accounts for self-destruction, colonisation and astrophysical destruction mechanisms of civilisations including gamma-ray bursts, type Ia and type II supernovae as well as radiation from the supermassive black hole. We adopt conservative estimates regarding the efficiency of such processes and find that astrophysical effects can influence the development of intelligent civilisations  and change the number of systems with such civilisations by roughly a factor of $2$; potentially more if the feedback is enhanced. Our results show that non-equilibrium evolution allows for solutions in-between extreme cases such as "rare Earth" or extreme colonisation, including scenarios with civilisation fractions between $10^{-2}$ and $10^{-7}$. These would imply still potentially large distances to the next such civilisations, particularly when persistence phenomena are being considered. As previous studies, we confirm that the main uncertainties are due to the lifetime of civilisations as well as the assumed rate of colonisation. For SETI-like studies, we believe that unbiased searches are needed considering both the possibilities that the next civilisations are nearby or potentially very far away.
\end{abstract}


\selfcitation{Schleicher DRG \& Bovino S (2022). The Fermi paradox: Impact of astrophysical processes and dynamical evolution. International Journal of Astrobiology. https://doi.org/10.1017/xxxxx}

\received{xx xxxx xxxx}

\revised{xx xxxx xxxx}

\accepted{xx xxxx xxxx}

\maketitle

\section{Introduction}
Our Universe is $13.76$~Gyrs old \cite{Komatsu2011}, and estimates suggest that our Galaxy could be colonised by advanced civilisations on the Fermi-Hart timescale \cite{Hart1975, Tipler1980} of \begin{equation}
	t_{FH}=10^6-10^8\mathrm{\ years},
\end{equation}
taking into account the size of the Milky Way and rough estimates for the travel time between stars. The fact that it did not happen or at least that we are not aware of any evidence for it is generally referred to as the Fermi paradox, expressed via the question "Where is everybody?" over a conversation at lunch. 

A very large range of possible solutions have been suggested for this problem (see \cite{Cirkovic2009} for a relatively recent review), which can be grouped into some typical classes of solutions: a) The "rare Earth"-type solutions going back to \cite{Ward2000}, which basically argues that the formation of intelligent, advanced civilisations requires so rare and specialized conditions that it could effectively happen only once in our Galaxy or even once in our Universe. b) Catastrophic solutions that basically suggest the self-destruction of advanced civilisations before they can start the colonisation process, considering options like Nuclear War, misuse of biotechnology, artificial intelligence or others \cite{Cirkovic2004b,Bostrom2008}. c) The possibility that advanced civilisations were actually here, but the expansion process rapidly moves forward and leaves behind an uncolonised wasteland \cite{Stull1979, Finney1985}. d) The class of solipsist solutions \cite{Sagan1983}, who suggest that the reason for the apparent absence of extraterrestrial civilisations is more related to the limits of our observations both of the Solar System as well as the Milky Way galaxy. A variant of this includes the zoo hypothesis \cite{Ball1973} which suggests that advanced extraterrestrial civilisations would rather avoid direct contact or visible manifestations of themselves, but rather observe us from the outside, for reasons of ethics, prudence or practicality \cite{Deardorff1987}.

\Fpagebreak

Of course, while explanations based on social dynamics of advanced civilisations can be a possibility, it is something out of the reach of practical investigations. On the other hand, the history of science on the long term has told us that our place in the Universe is not particularly special; we had to abandon the geocentric and the heliocentric theory, and modern cosmology suggests that the Universe is homogeneous and isotropic at least on large scales. In this light, it is worth repeating the argument given by \cite{Kinouchi2001}, which basically concerns independent civilisations on Earth. Very similar to the Fermi hypothesis, one could calculate the time scale it requires for a nation to establish contact with everybody on Earth, and independent of the details on how it is done, it surely would be feasible within less than $100$~years. The Brazilian government estimates that there are two hundred not contacted populations inside Amazonia,  and there are many lost islands in Indonesia, roughly 6000 of them inhabited but without contact to civilisation. If any member of such a not contacted population was to ask Fermi's question, they could easily conclude that no other human populations exist on Earth, since otherwise they should have been contacted by them. Other arguments to consider include considerations about the sustainability of an exponential expansion \cite{Haqq2009}, or potential problems due to the limited speed of light, potentially prohibiting efficient communication over very large scales. The Fermi paradox thus needs to be regarded with a grain of salt, to which many possible factors may contribute. 

Particularly, it is important to say that classic estimates based on e.g. the Drake equation \cite{Burchell2006} include, in addition to large uncertainties, many implicit assumptions, such as the suggestion that an equilibrium state has been effectively reached, which may or may not be the case, and some of the corresponding limitations were already pointed out by \cite{Cirkovic2004b}. Dynamical generalizations of the Drake equation have been explored e.g. by \cite{Panov2018}, and the age distribution of potential intelligent life in our Milky Way has been estimated by \cite{Legassick2015} taking the astrophysical formation of the Galaxy into account. More recently, a stochastic formulation of the Drake equation as a balance of birth-death processes was presented by \cite{Kipping2021}.

Such dynamical generalisations make sense, as the Universe and our Galaxy themselves are dynamically evolving, the number of habitable systems is growing over time, but there are also astrophysical processes that could have very destructive impacts on the development of complex forms of life, particularly at earlier times in the Universe, which can only be taken into account within dynamical models. Already \cite{Annis1999} suggested that astrophysical events like gamma-ray bursts could have a large destructive impact in our Galaxy, and potentially prohibit the formation of civilisations over very large scales. \cite{Galante2007} estimated that such events could even affect planetary systems more than $150$~kpc from the original event. \cite{Cirkovic2008} thus concluded that these or other events could lead to a global reset and in case such events were very common in the early Universe, may simply have prohibited the development of intelligent civilisations until recently. These models are nonetheless highly simplified, as gamma-ray bursts are highly beamed rather than spherical explosions, thus affecting only a smaller part of the galaxy.

In addition to gamma-ray bursts, there are other explosive events in our Galaxy, in particular the type Ia and type II supernovae, whose effects on astrobiology were already estimated e.g. by \cite{Gowanlock2011}. Particularly in the first two billion years, radiation from the activity of the supermassive black hole in our Galaxy could have affected the development of life, particularly in the inner region, the central few kiloparsecs \cite{Balbi2017}. 

In this paper, we aim to explore the population dynamics of advanced extraterrestrial civilisations, taking into account astrophysical constraints, the dynamical evolution within a non-equilibrium model, the impact of astrophysical events such as black hole activity, gamma-ray bursts and supernova explosions, as well as the possibility of colonisation and auto-destruction of civilisations via catastrophic events, which we parametrise due to the uncertainties. In the following, we will present first estimates based on the Drake equation, and then introduce a non-equilibrium model that includes the above-mentioned processes. Subsequently we present the results of the model and discuss the astrobiological implications.

\section{The Drake equation: first estimates}\label{Drake}

\begin{figure*}
    \centering
	\includegraphics[scale=0.3]{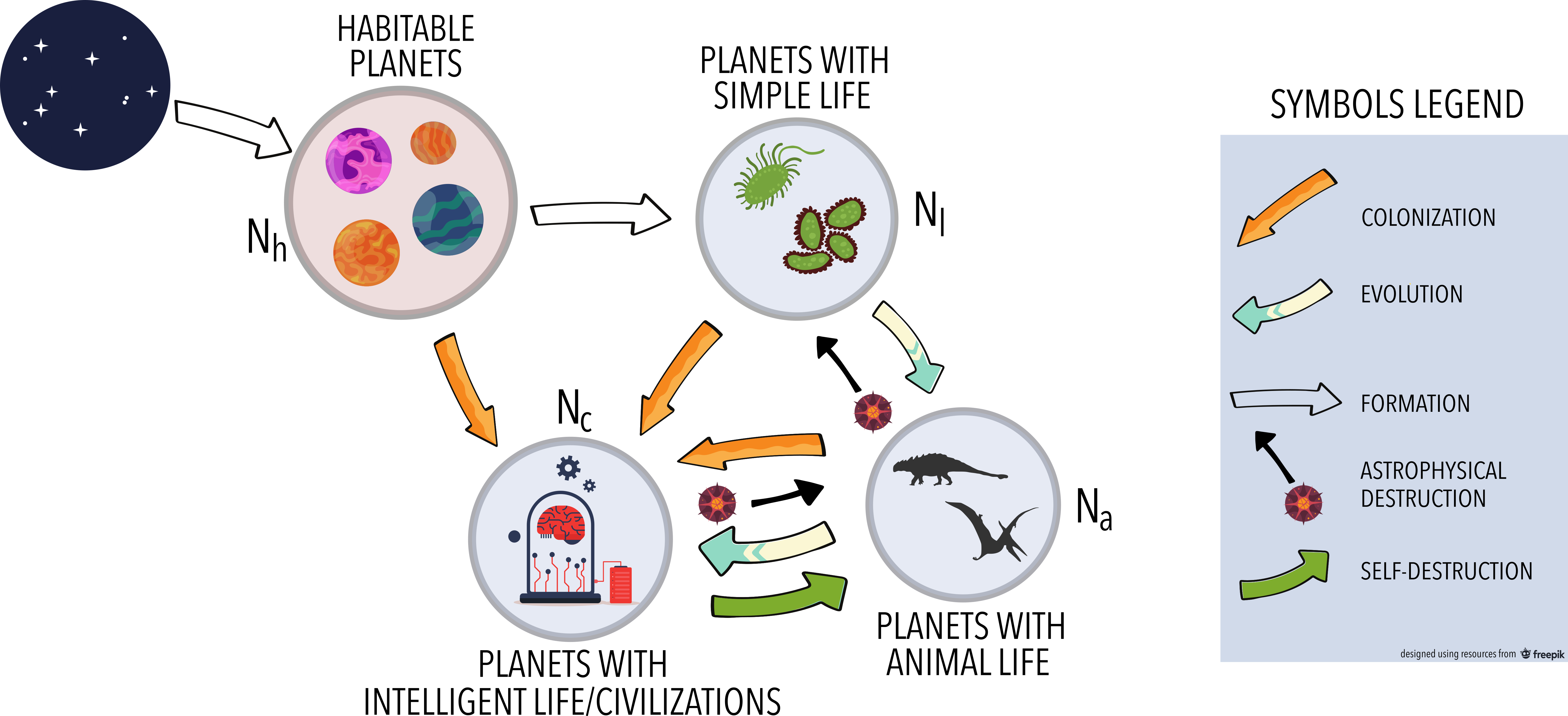}
	\caption{Illustration of the dynamical model described via Eqs.~\ref{Nh}-\ref{Nc}. A symbols legend is also reported for clarity.}\label{illu}
\end{figure*}

The Drake equation aims to estimate the number of communicating civilisations and is given, in its original form, as\begin{equation}
	N_{CC}=R_* f_p n_e f_lf_if_cL,
\end{equation}
where $R_*$ is the formation rate of stars in our galaxy averaged over all times, $f_p$ the fraction of stars that host planetary systems, $n_e$ the average number of planets per system suitable for life, $f_i$ is the fraction of such planets where intelligent life has developed, $f_c$ is the fraction of such planets where life has reached the communicative phase, and $L$ is the average duration of the communicative phase.

Some of these parameters are very well-known; {we have a star formation rate of $\sim1.9$~M$_\odot$~year$^{-1}$ \cite{Twarog1980,Meusinger1991, Chomiuk2011}. As shown by the work of \cite{Kroupa2003}, the initial mass function of stars is peaked in the low-mass regime, such that the average number of stars per one solar mass is about 2. This implies the formation of roughly 4 stars per year.} In the Milky Way galaxy, overall a rather quiescent evolution seems to have taken place over its history \cite{Kennicutt2012}. To reproduce the $\sim10^{11}$~stars in the Milky Way today, one needs to assume on average a rate $R_*\sim10$~year$^{-1}$ for the last $10$ billion years \cite{Prantzos2013}.

{We focus here predominantly on planets in the super-Earth regime with 2 up to 10 Earth masses, as defined by \cite{Stevens2013}. An upper limit on their abundance can be obtained from the analysis of \cite{Mayor2011}, who estimate the fraction of planets with less than 30 Earth masses as about $10\%$.} Another estimate has been made by  \cite{Selsis2007}, requiring stars to be sufficiently long-lived for the development of complex life forms (with main-sequence lifetimes larger than $4.5$~Gyr, and thus masses less than $1.1$~M$_\odot$) while having sufficiently high masses to have habitable zones outside the tidally locked regions, where always the same side of the planet would be exposed to radiation from the star. Based on these considerations, they arrived at a fraction of about $10\%$. We also assume here that within the habitable zone, we will only have about one planet per system. As these are independent requirements that need to be simultaneously fulfilled, we may then estimate $f_pn_e\sim0.1\cdot0.13=0.013$ \cite[see also][for a similar discussion]{Prantzos2013}.

From astrophysical considerations, we thus arrive at a rate $R_{\rm astro}=R_*f_pn_e\sim${0.052} habitable planets per year. We can further group the biological terms, $f_b=f_lf_i$, describing the fraction of those planets where intelligent life developed. The only reference system we can consider here is Earth, which has an approximate total age of $4.5$~Gyrs, while the first {hominids} appeared about $6$~million years ago. Based on these considerations, one could estimate $f_b\sim\frac{0.006}{4.5}\sim10^{-3}$, which might be considered as an upper limit. To estimate the fraction of planets where life has reached the communicative phase, again our only reference point is Earth, where radio signals exist since roughly the 1950s. {Based on this consideration, one might infer a possible minimum value of the corresponding fraction, of course under the assumption that typical life will behave similarly as on Earth, which would be $f_{\rm c,min}\sim\frac{70}{6\times10^6}\sim10^{-6}$. On the other hand, this fraction may also be considerably larger, and also the human society may keep communicating for a considerable amount of time. With the average duration of the communication phase $L$, we could make an improved estimate $f_c\sim \frac{L}{6\times10^6\mathrm{\ years}}$, implying $f_cL\sim\frac{L^2}{6\times10^6\mathrm{\ years}}$.}

Putting the numbers together, we thus obtain\begin{equation}
	N_{CC}\sim8.7\times10^{-12}\left( \frac{L}{\mathrm{years}} \right)^2.
\end{equation}
The number of communicating civilisations thus depends very strongly on their assumed lifetime, and of course it is still important to note that it would require a very advanced civilisation to detect the radio signals from Earth over a large distance. Nonetheless, for lifetimes of $L\sim10^6$~years, we could then expect $N_{CC}\sim${9} communicating civilisations in our Galaxy.

A very relevant uncertainty can be removed, however, if {we employ a modified version of the Drake equation where we aim to estimate the total number of civilisations, irrespective of whether they are communicating or not. We denote the number of total civilisations as $N_c$. It can be estimated by removing the factor $f_c$ from the Drake equation and by replacing the lifetime $L$ of the communicative phase with the total lifetime $L_{\rm tot}$ of an average civilisation. We obtain the following estimate:}
\begin{equation}
	N_{c}\sim5.2\times10^{-5}\left( \frac{L_{\rm tot}}{\mathrm{years}} \right).
\end{equation}
If we take $L_{\rm tot}\sim10^7$~years, which can be easily justified with the timescale since {hominids} are living on Earth, we arrive at $N_C\sim${520} civilisations. 

These estimates in particular do not yet take the possible process of colonisation into account. While our civilisation has not started such a process yet, it is worth noting that Earth formed only $4.5$~Gyrs ago, and other civilisations may have formed even Gyrs before ours. A more complete model thus needs to take this process into account. {We further point out another important limitation of the Drake equation: being tied to the star formation rate, it assumes civilisations to exist only on planets around recently formed stars. It is not obvious that this assumption is correct and even in the complete absence of current star formation, it would be conceivable that extraterrestrial civilisations exist on planets that formed billions of years ago. Another improvement of a dynamical framework is thus that it allows to take both the recently formed planets plus the already existing ones into account.}

\section{A non-equilibrium model}\label{model}

We now go beyond the Drake equation and present a dynamical non-equilibrium model which takes into account the process of colonisation as well as astrophysical destruction mechanisms of civilisations. Within this framework, we will consider the number of habitable planets $N_h$ {which provide conditions for life to develop though it has not developed yet.} $N_l$ {denotes} the number of systems where simple forms of life have developed, $N_a$ the number of systems including animal life forms on land and finally $N_c$ the number of systems with civilisations. {The total number of habitable planets at a given time is thus given as $N_h+N_l+N_a+N_c$, which is increasing over time due to ongoing star formation.} 

The evolution equations in our model are given as (see also Fig.~\ref{illu} for an illustration)
\begin{eqnarray}
	\dot{N}_h&=&R_{\rm astro}(t)-\epsilon_l N_h-\epsilon_{\rm col}N_cN_h/N,\label{Nh}\\
	\dot{N}_l&=&\epsilon_l N_h-\epsilon_a N_l+\epsilon_{\rm ast,a}(t)N_a-\epsilon_{\rm col}N_cN_l/N,\\
	\dot{N}_a&=&\epsilon_a N_l-\epsilon_c N_a+\epsilon_d N_c+\epsilon_{\rm ast,c}(t)N_c\nonumber\\
	&-&\epsilon_{\rm ast,a}(t)N_a-\epsilon_{\rm col}N_cN_a/N,\\
	\dot{N}_c&=&\epsilon_c N_a-\epsilon_dN_c+\epsilon_{\rm col}N_c(N_h+N_l+N_a)/N\nonumber\\
	&-&\epsilon_{\rm ast,c}(t)N_c,\label{Nc}
\end{eqnarray}
where $N=N_h+N_l+N_a+N_c$ is the total number of systems, $R_{\rm astro}(t)$ is the astrophysical formation rate of habitable systems  and $\epsilon_l$, $\epsilon_a$, $\epsilon_c$ $\epsilon_{\rm col}$, $\epsilon_d$, $\epsilon_{\rm ast,c}$ and $\epsilon_{\rm ast,a}$ denote various efficiency parameters that are introduced in further detail below.

First, we aim to derive a time-dependent model for the formation rate of habitable planets. As the evolution of star formation in the Milky Way overall was rather quiescent \cite[e.g.][]{Kennicutt2012}, we assume only a moderate time-dependence, following an exponential decay over a long timescale of $\sim5$~Gyrs. We choose our normalization such that over a timescale of $10$~Gyrs, the same number of systems will be produced with the time-averaged rate we adopted in the Drake equation above. We then have\begin{equation}
	R_{\rm astro}(t)=0.20\times10^9\mathrm{\ Gyr}\mathrm{\ \exp}\left( -\frac{t}{5\mathrm{\ Gyr}}. \right).
\end{equation}
Here $t$ is the time coordinate in our model, which is chosen such that at the present day we have $t=0$.

As initial conditions, we assume here that $N_h=N_l=N_a=N_c=0$, so all the systems form via the production term $R_{\rm astro}(t)$. $\epsilon_l$ is the efficiency parameter describing the transition from a habitable planet to a planet with simple forms of life (cells or bacteria), which we parametrise as $\epsilon_l=T_l^{-1}$. $T_l$ is the characteristic timescale for a simple form of life to develop. We adopt here  $T_l=10^9$~years, comparable to the time it took for the first life forms to develop on Earth. The term $\epsilon_a$ describes the transition from a planet with simple forms of life to planets with animal-type forms of life that live on land. We consider this as a particularly important intermediate stage that was a precondition for the subsequent development of civilisations. We adopt here $\epsilon_a=T_a^{-1}$, with $T_a\sim3$~Gyr considering the evolutionary time on Earth.

For the development of intelligent life and civilisations, similar as \cite{Annis1999}, we assume that roughly $0.1$~Gyr are required to develop sufficiently complex base classes of land-dwelling creatures, and subsequently another $0.1$~Gyr to develop intelligence once sufficiently complex base classes were being established. For the parameter $\epsilon_c$ describing the transition to intelligent life and civilisations, we thus adopt $\epsilon_c=T_c^{-1}$, with $T_c\sim0.2$~Gyr.

\subsection{Astrophysical destruction mechanisms}
We now consider astrophysical effects that can potentially destroy civilisations and even affect animal life. We parametrise the effect on civilisations via $\epsilon_{\rm ast,c}(t)=\epsilon_{\rm AGN}(t)+\epsilon_{gb}(t)+\epsilon_{\rm SNIa}(t)+\epsilon_{\rm SNII}(t)$, including feedback from the galactic supermassive black hole, gamma-ray bursts as well as type Ia and type II supernovae. The effect of such feedback on animal life is incorporated as well, which we parametrise via $\epsilon_{\rm ast,a}=a\epsilon_{\rm ast,c}$ with $a\sim0.25$. The parameter $a$ is of course very uncertain, though we consider here that more simple types of small animals should have higher chances to potentially survive catastrophic events.

For up to the first two billion years in our Galaxy, the supermassive black hole has likely been active, and we parametrise the effect of its feedback following \cite{Balbi2017}, who calculated that it would very severely affect planetary atmospheres and ecosystems up to $3$~kpc from the center of the Galaxy. Considering a typical size of $30$~kpc of the Galactic disk, the fraction of systems affected via the feedback thus amounts to $p_{\rm AGN}\sim0.01$. We adopt a characteristic timescale of $T_{\rm AGN}\sim10^{-3}$~Gyrs for the feedback events from the black hole, and thus arrive at an efficiency $\epsilon_{\rm AGN}=p_{\rm AGN}/T_{\rm AGN}$ during the first $2$~Gyrs, and subsequently $\epsilon_{\rm AGN}=0.$

For gamma-ray bursts, it is harder to make precise estimates, as their explosions are not spherical but highly beamed. In principle there are uncertainties given their absolute event rates, as they can only be detected from Earth if Earth falls into the beam of the gamma-ray burst. In case of smaller beams, the event rate is thus higher than typically estimated, and smaller for larger beams. Regarding the overall efficiency of the feedback, this uncertainty nevertheless is balanced out as a higher event rate would then correspond to an accordingly smaller part of the Galaxy affected by the event. For simplicity, what we adopt here is the estimate by \cite{Melott2004}, who assume that planetary systems would be strongly affected within $3$~kpc of the gamma-ray burst, thus corresponding to a fraction of $p_{gb}=0.01$ of the Galactic disk. Again following \cite{Melott2004}, we assume that currently in the Universe, such events occur roughly at a period of $P_\gamma=170$~Myrs. This rate should be expected to evolve over time, and we assume it to decay over a characteristic timescale of $t_\gamma\sim5$~Gyr, consistent with cosmological observations \cite{Bromm2002}. We thus arrive at the following event rate\begin{equation}
	\nu_\gamma(t)=\nu_0\mathrm{exp}\left( -\frac{t}{t_\gamma} \right)
\end{equation} 
with $\nu_0=1/P_\gamma$ and the corresponding efficiency parameter is thus given as\begin{equation}
	\epsilon_{gb}(t)=p_{gb}\nu_\gamma(t).
\end{equation}
For the rate of type Ia supernovae, we adopt the estimated rate of $\nu_{\rm SNIa}=4\times10^{-3}$~yr$^{-1}$ from \cite{Gowanlock2011} for the last billion years, consistent also with \cite{Meng2010}. The type Ia supernova rates are expected to evolve as the Galactic star formation rate, though with a delay time of $1$~Gyr due to stellar evolution effects. We assume here a similar decay law as for the gamma-ray bursts, though with a different frequency. For the type Ia supernovae, \cite{Gowanlock2011} derived a sterilization radius of 19~pc, corresponding to a fraction $p_{\rm SNIa}\sim4\times10^{-7}$ of the Galactic disk. Accounting for the delay time of $1$~Gyr, we write\begin{equation}
	\epsilon_{SNIa}(t)=p_{SNIa}\nu_{SNIa}\ \mathrm{exp}\left( -\frac{t-1\mathrm{\ Gyr}}{t_\gamma} \right).
\end{equation}
For the type II supernovae, we recall the sterilization radius derived by \cite{Gowanlock2011}, corresponding to $8$~pc or a fraction of $p_{\rm SNII}=0.008^2/30^2\sim7\times10^{-8}$ of the Galactic disk. For their event rate, we adopt a value of $\nu_{\rm SNII}=2\times10^{-2}$~yr$^{-1}$, which is in-between the \cite{Gowanlock2011} estimates.  We thus have\begin{equation}
	\epsilon_{\rm SNII}(t)=p_{\rm SNII}\nu_{\rm SNII}\ \mathrm{exp}\left( -\frac{t}{t_\gamma} \right),
\end{equation}
assuming the time dependence to be roughly the same as for the gamma-ray bursts. 

\subsection{Colonisation}
We finally arrive at the uncertainties that are set by the potential behavior or evolution of intelligent civilisations themselves, which we will regard here as essentially unknown, and we will aim to avoid here any \textit{a priori} assumptions to the degree possible. Assuming that advanced civilisations will aim to survive catastrophic destruction events of their host stellar system due to the finite lifetime of the star, one could assume that at least a few colonisation events should occur over the characteristic stellar evolution time of $\sim10$~Gyrs in case of a solar mass star, and we parametrise here the colonisation efficiency as
\begin{equation}
	\epsilon_{\rm col}=\frac{N_{\rm new}}{10\mathrm{\ Gyr}},
\end{equation}
where $N_{\rm new}$ the number of new colonies a typical civilisation would produce over a time of $10$~Gyrs (assuming they exist long enough). Depending on the typical behavior of such civilisations, of course $N_{\rm new}$ could be considerably larger than a factor of a few, and we will later explore the dependence on this parameter. The other relevant and basically unknown factor is the typical lifetime of a civilisation. On Earth, the first hominids appeared roughly $6$~million years ago, which sets a possible lower limit for the existence of an intelligent species (though most of that time had not been spent at a technologically advanced stage). It may be the case that the lifetime of a civilisation becomes shorter once a significant level of technological advance has been reached, but is also not completely obvious. We therefore will regard it as an unknown parameter as well, and parametrise it as $\epsilon_d=T_d^{-1}$.

\section{Results}\label{results}

\begin{figure}
	\includegraphics[scale=0.5]{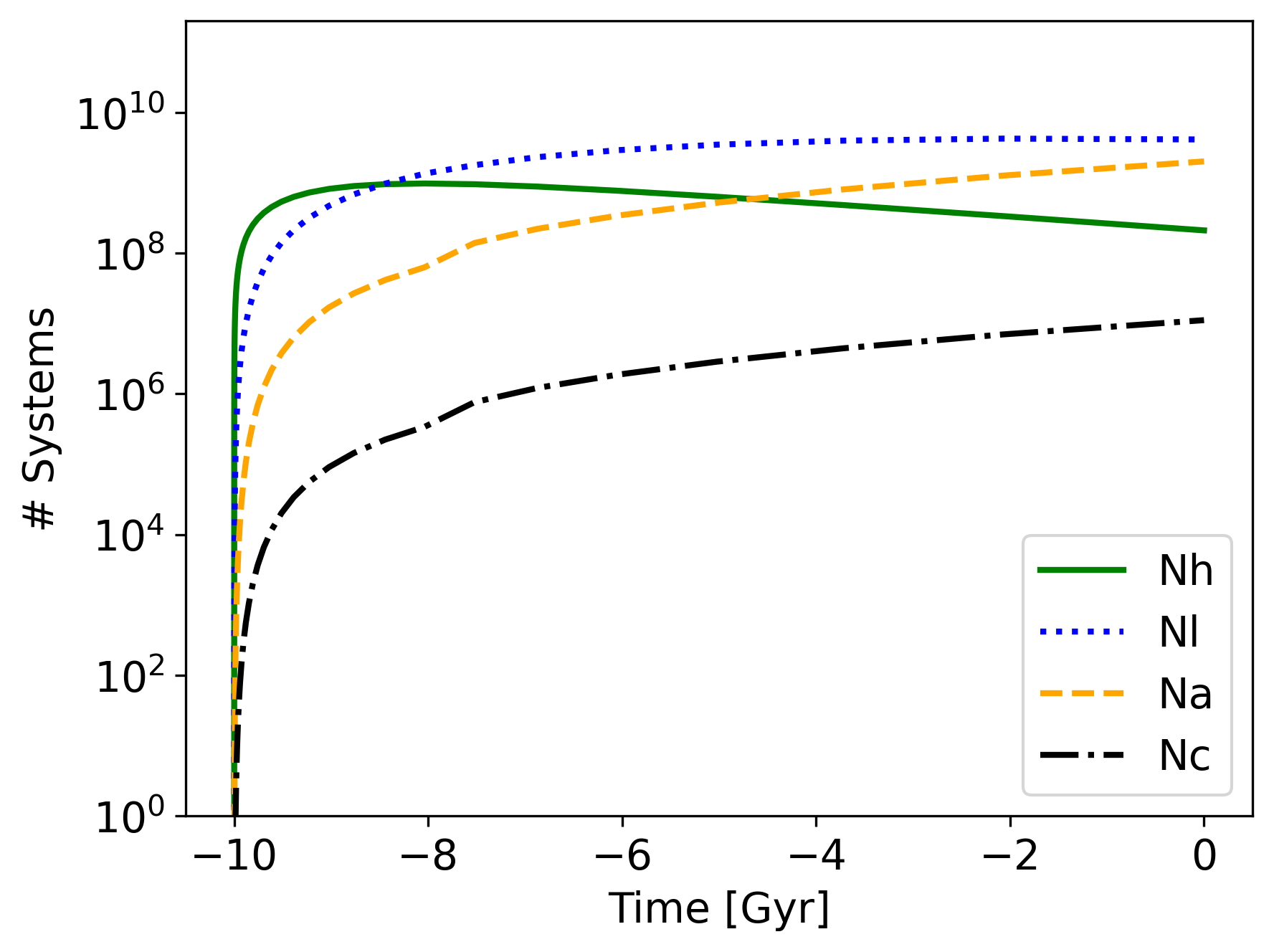}
	\caption{Our reference model with $N_h$, $N_l$, $N_a$ and $N_c$ as a function of time.\label{reference}}
\end{figure}

We solve the set of first order ordinary differential equations developed 
above by employing the \textsc{scipy}'s \textsc{solve\_ivp} Python function, an implicit multi-step variable-order BDF solver.
The system of equations is integrated from time $t=-10$~Gyrs until $t=0$, in a logarithmically spaced grid of $100$ points\footnote{Relative and absolute tolerances are set to \textsc{rtol=10$^{-12}$} and \textsc{atol=10$^{-40}$}, respectively.}. We recall that we only model here the habitable systems, so roughly a fraction of $f_pn_c\sim0.013$ of all systems, as discussed in our estimates regarding the Drake equation. As a reference model, we adopt the parameters $T_d=10^{-3}$~Gyrs and $N_{\rm new}=10^3$ to show the results of the model in a specific case, and we also show how it is sensitive to changes and variations in these parameters. We summarize the main input parameters for our reference model in Table~\ref{parameters}.

\begin{table}[]
    \centering
    \begin{tabular}{ccc}
    parameter     &  value & meaning \\ \hline
    $T_l$     &    $1$~Gyr & time to form simple life\\
    $T_a$ & $3$~Gyr & time to form animal life\\
    $T_c$ & $0.2$~Gyr & time to form civilisations \\
    $T_d$ & $10^{-3}$~Gyr & lifetime of a civilisation \\
    $N_{\rm new}$ & 1000 &  colonisation events per\\
    & & civilisation over $10$~Gyr\\
    $a$ & 0.25 & astrophysical feedback efficiency \\
    & & for animal life relative to civilisations \\ \hline
    \end{tabular}
    \caption{Input parameters for our reference model.}
    \label{parameters}
\end{table}

Within our reference model (see Fig.~\ref{reference}), in general a significant number of planetary systems builds up within the first billion years, where the formation rate of these systems is enhanced by a factor $e^2\sim7.3$, allowing to form roughly the same number of systems as in the subsequent $9$~Gyrs. These systems are habitable and so simple life starts to develop, and after roughly $2$~Gyrs there are more systems with simple life than without. After roughly $5$~Gyrs, the number of systems with animal life is larger than the number of systems with simple life. The number of civilisations grows steadily, and reaches about $10^7$ after $10$~Gyrs, thus corresponding to a fraction of $1\%$ of all habitable planets and $\sim0.01\%$ of all planetary systems. {The number of civilisations is thus considerably larger than based on the estimate in the Drake equation. The main reason is that the Drake equation only considers very recently formed stellar systems, as it is tied to the star formation rate, and with a lifetime of the civilisations of the order of one million years, the time they stay in that phase is short compared to the age of the Universe. The dynamical model considered here on the other hand allows for the development of civilisations on all of the habitable systems, including those that previously harbored a civilisation that subsequently disappeared. Further, also} the process of colonisation is taken into account, though still at a moderate level with $N_{\rm new}=1000$, which basically does not yet affect the evolution very much. An important effect that we see is that the number of civilisations keeps monotonically increasing and an equilibrium state has not been reached even after $10$~Gyrs{, and it is also not expected due to the continuously increasing number of available systems}. This underlines in principle the importance of a non-equilibrium treatment for the modeling of such populations.

\begin{figure}
	\includegraphics[scale=0.5]{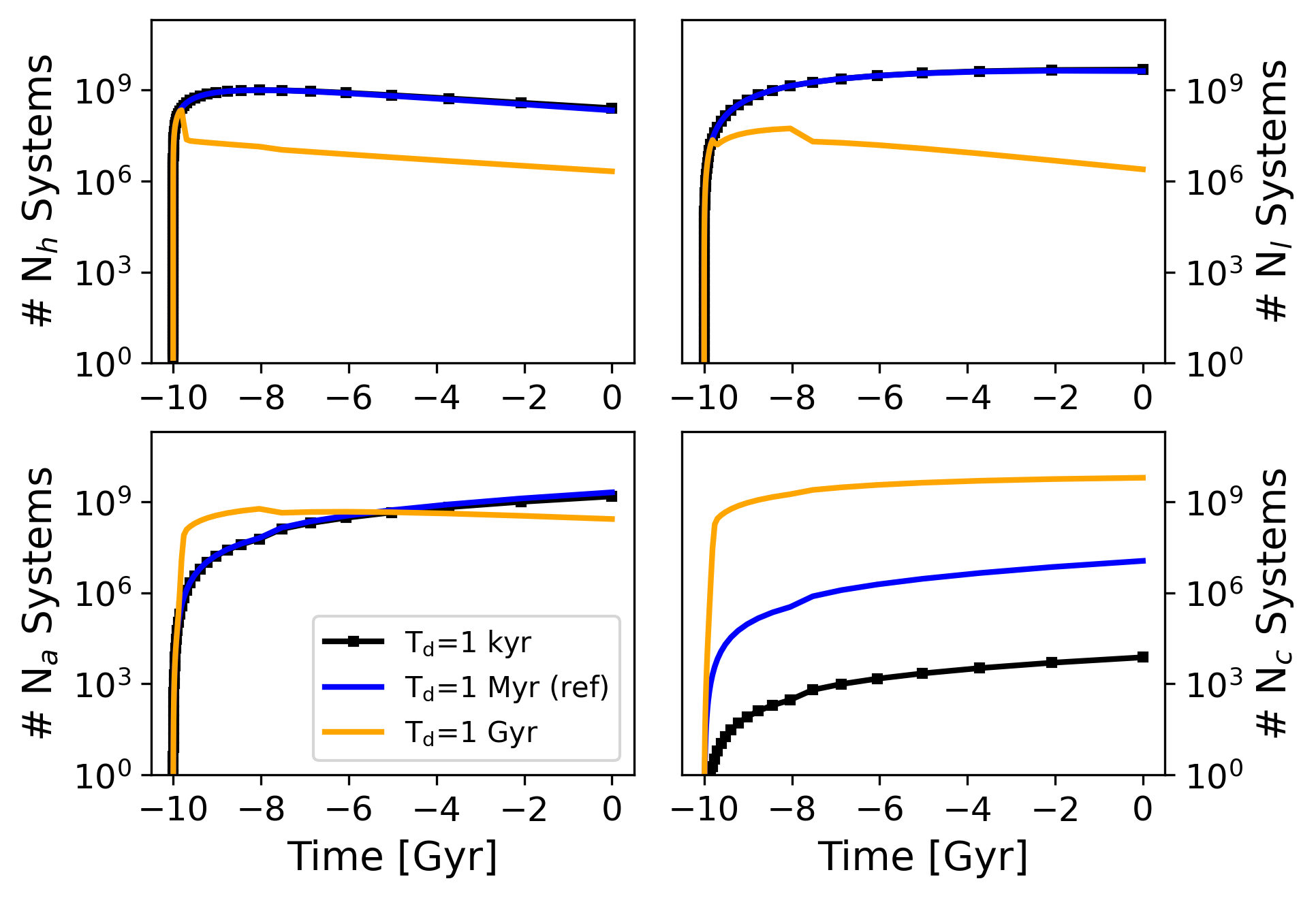}
	\caption{We show here the results of our reference model, varying the characteristic lifetime of a civilisation considering $T_d=1$~kyr, $1$~Myr and $1$~Gyr. The top left panel shows he number of habitable systems as a function of time, the top right one the number of systems which have developed simple forms of life, the bottom left one systems with animal life on land, and the bottom right one systems that host civilisations.\label{Td}}
\end{figure}

We now focus on the effect of the parameter $T_d$, as a potentially short lifetime of a civilisation is often considered as a possible solution towards the Fermi paradox, and we show what happens when this parameter is varied in Fig.~\ref{Td}. As expected, we find it to significantly affect the number of systems with civilisations, which changes essentially by the same factor. It is interesting to note that even in the extreme case of $T_d=1$~kyr, still roughly $10^4$ civilisations are left, while in case of $T_d\sim1$~Myr, almost all of the systems include civilisations, with the fraction of systems that harbors only animal life decreasing by more than an order of magnitude at the end of the evolution.

\begin{figure}
	\includegraphics[scale=0.5]{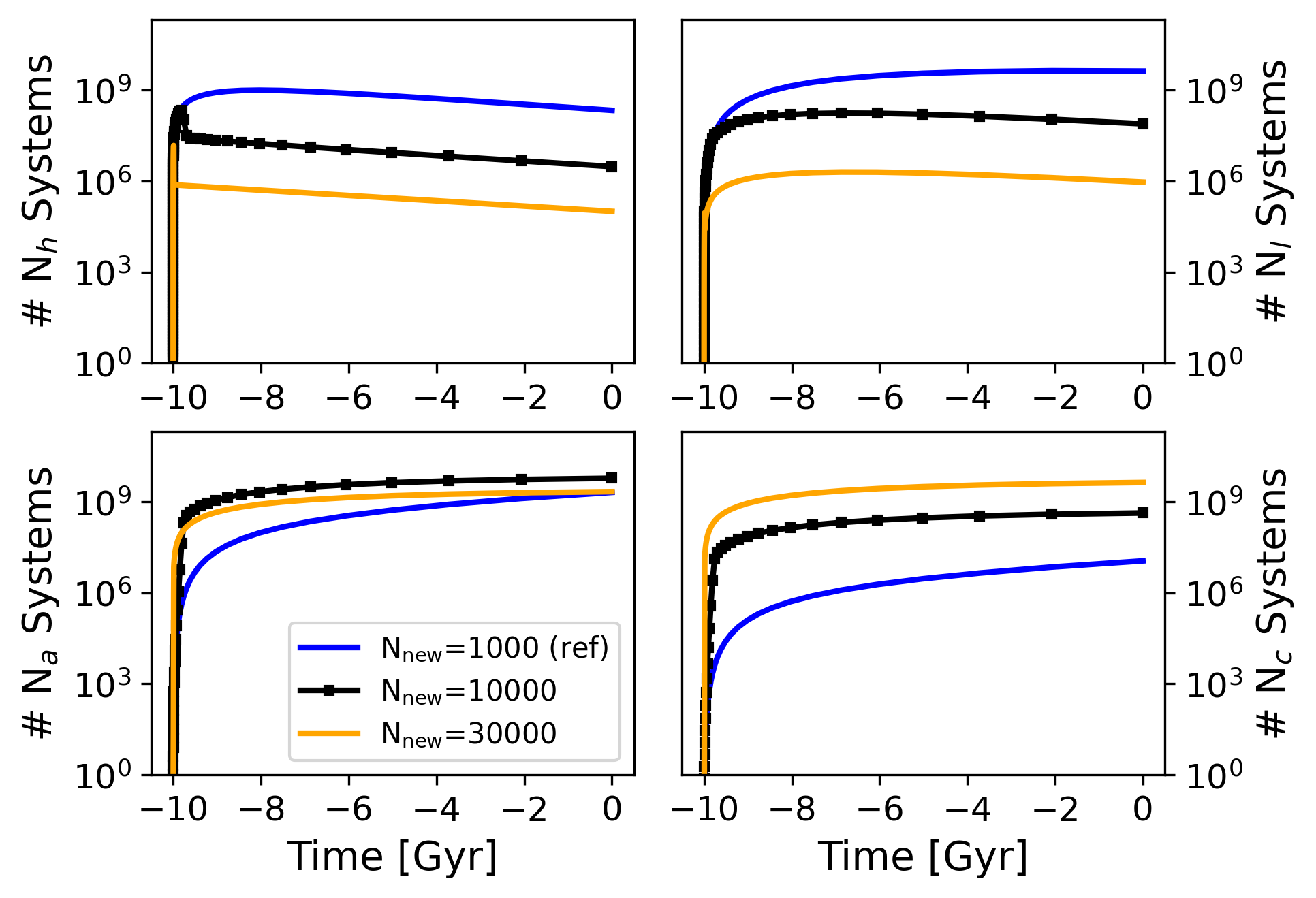}
	\caption{We show here the results of our reference model, varying the characteristic lifetime of a civilisation considering $N_{\rm new}=10^3, 10^4$ and $3\times10^4$. The top left panel shows he number of habitable systems as a function of time, the top right one the number of systems which have developed simple forms of life, the bottom left one systems with animal life on land, and the bottom right one systems that host civilisations.\label{Nnew}}
\end{figure}

We next consider the effect of varying the parameter $N_{\rm new}$, i.e. the number of colonies a civilisation would create over a timescale of $10$~Gyrs. This in principle tells us already that this effect will only be important if the lifetime of a civilisation will be sufficiently large, i.e. we need to have \begin{equation}
	\frac{10\mathrm{\ Gyr}}{N_{\rm new}}<T_d
\end{equation}
for colonisation to be efficient. We will consider here parameters $N_{\rm new}=10^3, 10^4$ and $3\times10^4$ (see Fig.~\ref{Nnew}). The first case is our reference scenario where we obtain about $10^7$ systems with civilisations, and we have $10$~Gyr/$N_{\rm new}>T_d$, so the process as noted above is not yet efficient. For $N_{\rm new}=10^4$, we are at the limiting case where $10$~Gyr/$N_{\rm new}=T_d$. At this point the colonisation process contributes in a relevant way to form civilisations, and $N_c$ increases by roughly an order of magnitude (as essentially the death of civilisations is compensated for). With $N_{\rm new}=3\times10^4$, we are at $N_c>10^9$, so at this point almost all habitable planets have colonies. A further increase of $N_{\rm new}$ would thus not make a relevant difference at the current time in the Universe, and in principle not even for most of the past, as indeed a high level of colonisation is then already reached after a few Gyrs.

\begin{figure}
	\includegraphics[scale=0.5]{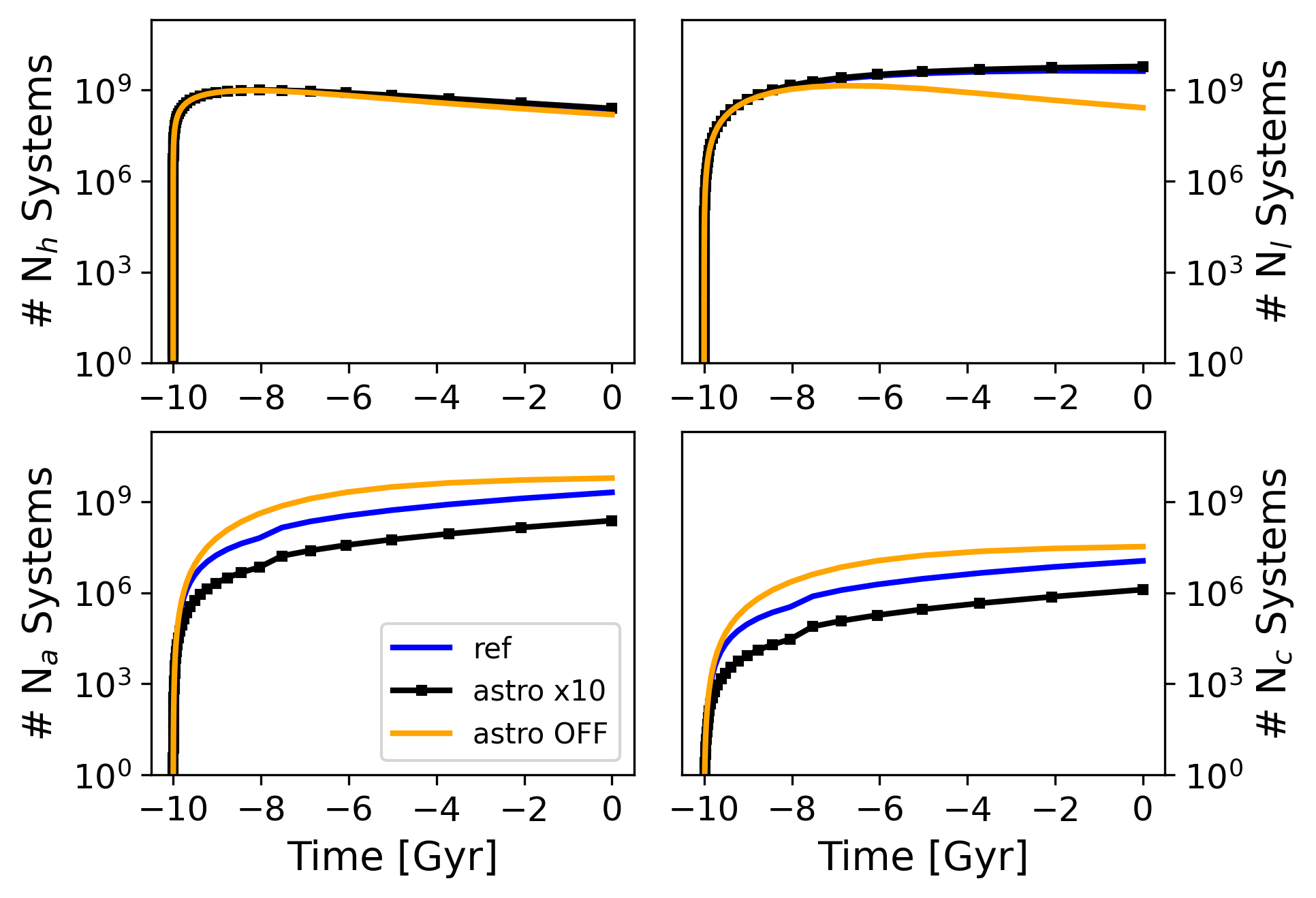}
	\caption{We show here the results of our reference model, comparing the results with and without astrophysical destruction mechanisms, as well as a case where the astrophysical destruction is enhanced by a factor of $10$. The top left panel shows he number of habitable systems as a function of time, the top right one the number of systems which have developed simple forms of life, the bottom left one systems with animal life on land, and the bottom right one systems that host civilisations.\label{astro}}
\end{figure}

An important part of our assessment further concerns the impact of astrophysical feedback mechanisms, such as radiation from the supermassive black hole, gamma-ray bursts as well as type Ia and type II supernovae on the evolution of civilisations and the model in general. For this purpose, we show our models with the astrophysical feedback implementation, without the feedback as well as a case where the feedback is artificially enhanced by a factor of $10$ in Fig.~\ref{astro}. The latter is done to account for potential uncertainties, both regarding the astrophysical parameters themselves, but also due to the uncertainties in their impact on planetary systems. It is probably fair to say that we adopted here somewhat conservative estimates of the impact of astrophysical effects. 

In spite of the uncertainties, one can probably say that these parameters are nonetheless better constrained than $T_d$ or $N_{\rm new}$, so it seems appropriate to vary them to a lesser degree. In spite of this, we find that the astrophysical feedback influences the development of civilisations by a relevant degree. In our reference case as before, we find $N_c\sim10^7$, which increases by roughly a factor of $3$ if feedback is being switched off. The increase is even larger at earlier times where the frequency of supernovae and gamma-ray bursts is still higher and the feedback thus more efficient. We find here that particularly the type Ia supernovae and next the type II supernovae are making a difference; gamma-ray bursts affect a larger fraction of the galaxy but nonetheless are much less frequent and thus overall less important. On the other hand, the feedback by the supermassive black {hole} matters only in the first two Gyrs where the civilisations in general are still building up and the feedback overall is less relevant. We note that, in the case we artificially increase astrophysical feedback by a factor of $10$, the number of civilisations decreases by a corresponding factor.

\begin{figure}
	\includegraphics[scale=0.5]{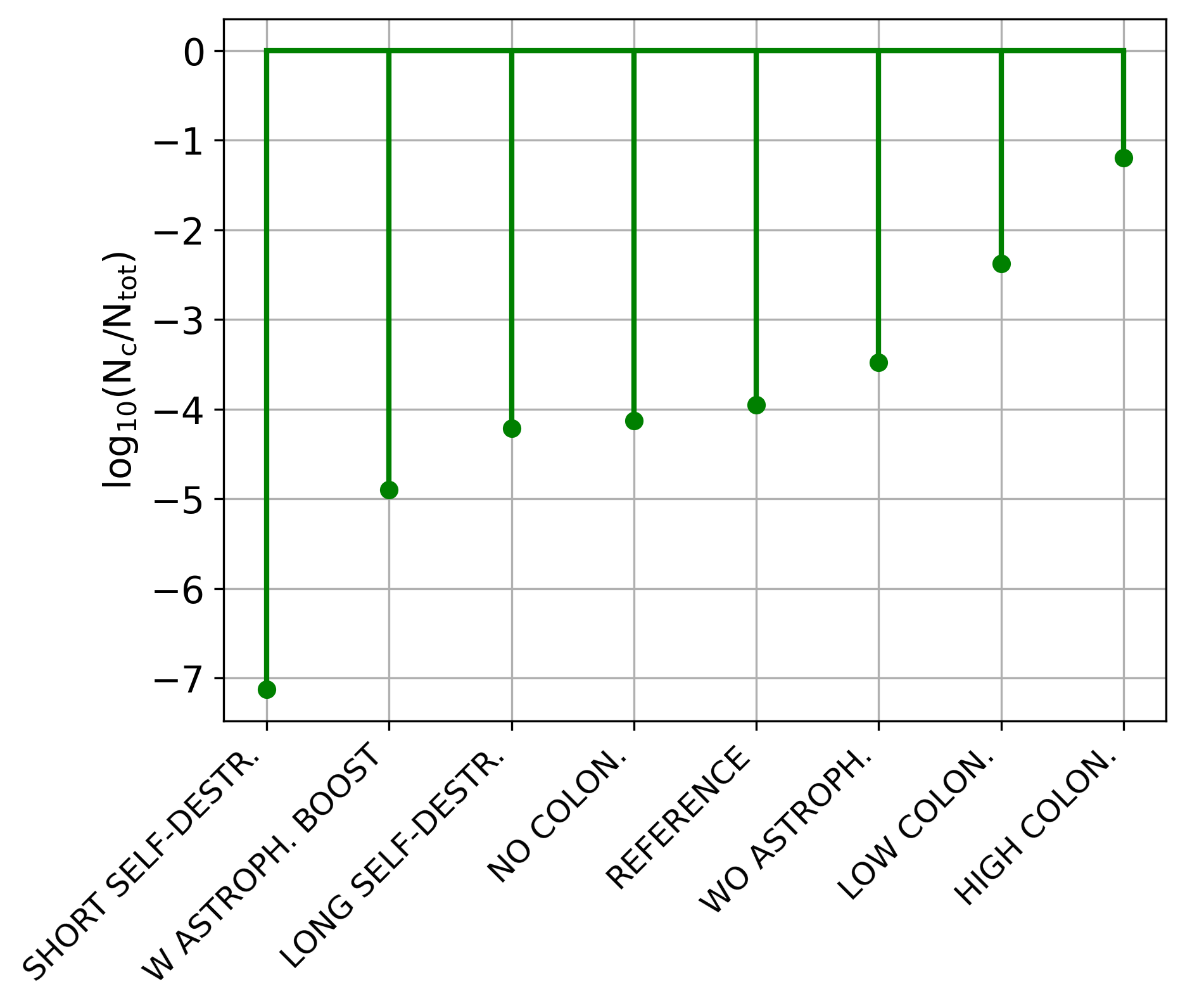}
	\caption{A summary of all the cases considered here, with the corresponding fractions of systems with civilisations that were obtained (relative to the total number of planetary systems).\label{summary}}
\end{figure}

For a comparison of the different effects considered here, we provide a summary plot in Fig.~\ref{summary}, showing the fraction of civilisations (with respect to all planetary systems, i.e. including the factor $f_pn_c$), to show how the latter varies under different conditions. Particularly, we find that a short lifetime of civilisations is most effective in suppressing their abundance, but also enhanced astrophysical destruction can bring down the fraction of civilisations. The efficiency of colonisation is very uncertain but can bring up the number of civilisations by a very high degree. Astrophysical feedback in general affects the number of civilisations by roughly a factor of $3$.

\section{Discussion and conclusions}\label{discussion}

We have developed and presented a non-equilibrium model that describes the emergence of civilisations, the process of colonisation as well as their destruction via astrophysical feedback mechanisms, including radiation from the central black hole in our Galaxy, via gamma-ray bursts and via type Ia and type II supernovae. The need for such dynamical modeling and a departure from the Drake equation has been pointed out in early works, including the potentially destructive effect of astrophysical processes \cite[e.g.][]{Annis1999, Cirkovic2004b, Galante2007, Cirkovic2008, Panov2018}. In principle we find the effect of astrophysical events to be less destructive as previously postulated in some studies, but nevertheless relevant, decreasing the number of civilisations by roughly a factor of $2$ and even more strongly at earlier times in the Universe. We have adopted here rather conservative estimates for the effect of astrophysical feedback, but also explored a case where it is enhanced by roughly a factor of $10$ to account for the uncertainties. It is of course in general very important to further study the astrobiological implications of such events, as we have indications that these may have influenced Earth at earlier times, and it is also important to consider the mitigation of such corresponding threats on the longer term \cite{Cirkovic2016}. In the context of the Fermi paradox, it currently appears that the corresponding effects are not strong enough to resolve it, but nonetheless they are a relevant factor for the modeling of the development of intelligent civilisations.

The two most uncertain parameters in the modeling of such civilisations are still their typical lifetimes as well as their colonisation rates, which potentially can vary by orders of magnitude. It is in principle thus unsurprising that they can lead to a similarly large variation in the results. What distinguishes our model from more simplified estimates is that it allows to treat and identify some intermediate cases, where small but nonetheless relevant fractions of intelligent civilisations exist within the Milky Way galaxy, and we have shown several cases here with outcomes in the range of $10^3$ up to $10^7$, which correspond to fractions of $10^{-8}$ up to $10^{-4}$ of all planetary systems. These solutions are interesting as they are in-between the extreme cases of the "rare Earth" solutions with effectively one civilisation per galaxy, or the Fermi paradox where almost every habitable system should be colonised. The fractions are also low enough to still imply a significant distance of any such civilisations from Earth, particularly if additional effects such as persistence are taken into account \cite{Kinouchi2001}. {Even in the regime where colonisation is not yet important, i.e. where the colonisation rate is lower than the death rate of civilisations, the estimates from our dynamical model imply much larger numbers of civilisations as compared to the Drake equation. This is due to the fact that in our dynamical model, these can form on all habitable stellar systems, while the estimates of the Drake equation are tied to the star formation rate and only consider the recently formed stars. It further implicitly assumes that each stellar system will develop a civilisation only once, while in our model it is feasible that a civilisation on a system develops, dies and later on a new one forms. The number of civilisations overall is continuously increasing, essentially due to the continuous increase of habitable systems.} 

We note also some of the simplifications we had to make; particularly the timescales for the development of simple life, animal life and civilisations were only motivated from the development on Earth and are very approximate; unfortunately so far we do not have another reference point to assess i.e. the possible variation in these parameters. We also note that the development of animal or intelligent life may not necessarily be restricted to super-Earth type planets as assumed here, and other channels for the development of such lifeforms may exist for instance in the moons of giant or rogue planets as well \cite{Avila2021}. The purpose of our work however was to provide a simple model and also to avoid speculation as much as possible, and thus to be rather conservative in our estimates.

In the context of SETI-like searches, of course one needs to obtain estimates not only for the number of civilisations, but also the number of communicating civilisations, which adds another degree of uncertainty, as we also do not know whether advanced civilisations will keep communicating or whether they might stop doing that, for reasons of security, energy efficiency or other types of considerations. Our results for the number of civilisations can of course be rescaled with any factor $f_c$ that appears well-motivated in a given context, though overall remains unknown. 

Regarding SETI-like projects, it is difficult to derive firm conclusions, and we believe that such projects should be pursued with an open mind, i.e. avoiding a priori assumptions on where other civilisations might be. Important discoveries can include the unexpected and a theory can only be validated or disproven via actual investigation. This should happen with as little bias as possible, considering both the possibilities of other civilisations (or their precursors) being close or distant.

\ack[Acknowledgement]{DRGS and SB gratefully acknowledge support by the ANID BASAL projects ACE210002 and FB210003, as well as via the Millenium Nucleus NCN19-058 (TITANs). DRGS thanks for funding via Fondecyt Regular (project code 1201280). SB is financially supported by ANID Fondecyt Regular (project code 1220033).}



\end{document}